\documentstyle[12pt,epsf]{article}

\begin{document}

\baselineskip=18.8pt plus 0.2pt minus 0.1pt

\makeatletter

\renewcommand{\thefootnote}{\fnsymbol{footnote}}
\newcommand{\beq}{\begin{equation}}
\newcommand{\eeq}{\end{equation}}
\newcommand{\bea}{\begin{eqnarray}}
\newcommand{\eea}{\end{eqnarray}}
\newcommand{\nn}{\nonumber\\}
\newcommand{\hs}[1]{\hspace{#1}}
\newcommand{\vs}[1]{\vspace{#1}}
\newcommand{\Half}{\frac{1}{2}}
\newcommand{\p}{\partial}
\newcommand{\ol}{\overline}
\newcommand{\wt}[1]{\widetilde{#1}}
\newcommand{\ap}{\alpha'}
\newcommand{\bra}[1]{\left\langle  #1 \right\vert }
\newcommand{\ket}[1]{\left\vert #1 \right\rangle }
\newcommand{\vev}[1]{\left\langle  #1 \right\rangle }

\newcommand{\ul}[1]{\underline{#1}}
\newcommand{\tr}{\mbox{tr}}
\newcommand{\ishibashi}[1]{\left\vert #1 \right\rangle\rangle }

\makeatother

\begin{titlepage}
\title{
\hfill\parbox{4cm}
{\normalsize KIAS-P02037\\{\tt hep-th/0206131}}\\
\vspace{1cm}
D-branes in NSNS and RR pp-wave backgrounds and S-duality
}
\author{Yoji Michishita
\thanks{
{\tt michishi@kias.re.kr}
}
\\[7pt]
{\it School of Physics, Korea Institute for Advanced Study}\\
{\it 207-43, Cheongryangri, Dongdaemun, Seoul, 130-012, Korea}
}

\date{\normalsize June, 2002}
\maketitle
\thispagestyle{empty}

\begin{abstract}
\normalsize
We investigate boundary conditions for open strings in NSNS and RR pp-wave
backgrounds constructed by Russo and Tseytlin, which are S-dual to 
each other. We show that if we do not turn on any boundary term 
(i.e. gauge field), D-branes in the RR background cannot move away from 
the origin in most cases, while those in the NSNS background can move 
anywhere. We construct RR counterparts of D3-branes in the NSNS background
as D3-branes with gauge fluxes and show that indeed they can move anywhere, 
in accord with S-duality. 
\end{abstract}

\end{titlepage}

\section{Introduction}

pp-wave backgrounds are rare examples that 
worldsheet theory of fundamental strings in them
is exactly solvable in spite of the presence of nonzero RR flux \cite{m,mt}.
In light cone gauge worldsheet action of fundamental strings in 
Green-Schwarz(GS) formalism is quadratic and solvable.
Closed string mass spectrum is that of a collection of 
infinite number of harmonic oscillators with mode numbers shifted by 
mass terms.

pp-wave backgrounds can be derived from AdS solutions by taking 
Penrose limit\cite{bfhp}. Limit version of AdS/CFT correspondence has been 
proposed in \cite{bmn}.

It is very important to investigate open strings and D-branes 
in these backgrounds for understanding string dynamics in RR backgrounds
and AdS/CFT correspondence. Some such attempts have been done in 
\cite{bp,dp,bgg} 
for pp-wave from $AdS_5\times S^5$. See also \cite{ch,chl,lp,bgmnn,bhln}.
Main results of \cite{bp,dp,bgg} are that half 
BPS D-branes must sit at the origin, 
and less supersymmetric configurations are inconsistent.

In this paper we investigate open string boundary conditions in 
NSNS and RR pp-wave backgrounds given in \cite{rt}. 
(Some analysis has been already done in \cite{kns}. Some supergravity 
solutions have been constructed in \cite{kns,ak}.) 
These two backgrounds can be obtained by taking limits of 
$AdS_3\times S^3\times T^4$, and S-dual to each other. BPS brane spectra
are expected to respect S-duality. We will see that indeed BPS 
D3-brane spectra coincides in an interesting way. 

This paper is organized as follows.
In section 2 we analyze boundary conditions of open strings 
in the RR background without introducing any boundary term. 
In particular we consider the conditions 
which preserve half of the dynamical supersymmetry.
We will find that in most cases D-branes must sit at the origin, 
similarly to the result in \cite{bp,dp}.

In section 3 we repeat the same analysis for the NSNS background.
We will find that BPS configurations are almost the same ones as 
in the flat background. In particular, D-branes can be anywhere.

Since the NSNS background has nonzero NSNS 2-form, 
it is D3-branes with nonzero gauge fields that are S-dual to 
D3-branes in section 3. 
In section 4 we construct such D3-branes in the RR background and 
show that indeed they can move anywhere. 
(Similar phenomenon in pp-wave background originating from
$AdS_5\times S^5$ has been reported in \cite{st}.)
Furthermore low-lying open string spectra on BPS D3-branes 
coincide with those of section 3. 
Section 5 contains conclusions.

\section{D-branes in RR pp-wave background}
In this section we consider open strings in the following pp-wave 
background with the RR 3-form flux\cite{rt}:
\bea
ds^2 & = & dudv-f^2x^ix^idu^2+dx^idx^i+dx^adx^a, \nn
C_{ij} & = & -{\cal F}_{ij}u,
\eea
where $C_{ij}$ is the RR 2-form, ${\cal F}_{i_1j_1}=2f\epsilon_{i_1j_1}$,
${\cal F}_{i_2j_2}=2f\epsilon_{i_2j_2}$, 
$i,j=1,2,3,4$, $i_1,j_1=1,2$, $i_2,j_2=3,4$ and $a=5,6,7,8$.
It is shown in \cite{rt} that this background can be obtained as a limit of
$AdS_3\times S^3\times T^4$. $x^a$ correspond to $T^4$.

The worldsheet action of fundamental strings in GS
formalism is
\bea
L & = & \frac{1}{\pi\ap}\Bigg[ \Half\p_+u\p_-v+\Half\p_+v\p_-u
-m^2x^ix^i
+\p_+x^i\p_-x^i+\p_+x^a\p_-x^a \nn
 & & +iS_R\p_+S_R+iS_L\p_-S_L+2imS_L\gamma^1\gamma^2R_+S_R \Bigg],
\eea
where we have taken the light cone gauge, in which this action 
is quadratic and therefore exactly solvable, $u=2\ap p^u\tau$
and $m=\ap p^uf$.
$\gamma^i$ are SO(8) $8\times 8$ real gamma matrices\footnote{
For an explicit expression, see page 288 of \cite{gsw}.} and
$R_\pm=\Half(1\mp\gamma^1\gamma^2\gamma^3\gamma^4)$.

We decompose $S_{R,L}$ into 4-component spinors $s_{R,L}$ 
and $\hat{s}_{R,L}$:
\beq
S_{R,L}=\left(\begin{array}{c}
s_{R,L} \\
\hat{s}_{R,L}
\end{array}\right),
\label{decomp}
\eeq
\beq
R_+ S_{R,L}=\left(\begin{array}{c}
s_{R,L} \\
0
\end{array}\right),\quad
R_- S_{R,L}=\left(\begin{array}{c}
0 \\
\hat{s}_{R,L}
\end{array}\right),\quad
\gamma^{12}S_{R,L}=-\left(\begin{array}{c}
\Lambda s_{R,L} \\
\hat{\Lambda}\hat{s}_{R,L}
\end{array}\right),\quad
\eeq
where $\Lambda$ and $\hat{\Lambda}$ are $4\times 4$ antisymmetric 
matrices with $\Lambda^2=\hat{\Lambda}^2=-1$.
We can see only $s_{R,L}$ have the mass term.
Since analysis for $x^a$ and $\hat{s}_{R,L}$ is trivial, we will often omit
them below.

This action has two types of supersymmetry i.e. dynamical 
supersymmetry and kinematical supersymmetry. They are given as follows.
\begin{itemize}
\item dynamical supersymmetry
\bea
\delta x^i & = & i\epsilon_R\gamma^iR_+S_R+i\epsilon_L\gamma^iR_+S_L, \\
\delta x^a & = & i\epsilon_R\gamma^aR_-S_R+i\epsilon_L\gamma^aR_-S_L, \\
\delta S_R & = & -\p_-x^i\epsilon_R\gamma^iR_+-\p_-x^a\epsilon_R\gamma^aR_-
 -mx^i\epsilon_L\gamma^i\gamma^1\gamma^2R_+, 
\label{dsrrsr}\\
\delta S_L & = & -\p_+x^i\epsilon_L\gamma^iR_+-\p_+x^a\epsilon_L\gamma^aR_-
 -mx^i\epsilon_R\gamma^i\gamma^1\gamma^2R_+,
\label{dsrrsl}
\eea
where $R_+\epsilon_{L,R}=0$ i.e. $\epsilon_{L,R}$ have 4 nonzero components.
\item kinematical supersymmetry
\bea
\tilde{\delta} S_R & = & e^{-2m\gamma^1\gamma^2R_+\tau}\tilde{\epsilon}_R
+e^{2m\gamma^1\gamma^2R_+\tau}\tilde{\epsilon}_L,\nn
\tilde{\delta} S_L & = & e^{-2m\gamma^1\gamma^2R_+\tau}\tilde{\epsilon}_R
-e^{2m\gamma^1\gamma^2R_+\tau}\tilde{\epsilon}_L.
\eea
\end{itemize}

In this section we do not introduce nonzero gauge field on D-branes.
Then variation of the action is 
\bea
\delta S & = & 
\frac{1}{\pi\ap}\int d^2\sigma \Bigg[ -2\delta x^i(\p_+\p_-x^i
-m^2x^i)-2\delta x^a\p_+\p_-x^a \nn
& & 
+i\delta s_R(\p_+s_R-m\Lambda s_L)
+i\delta s_L(\p_+s_L-m\Lambda s_R)
+i\delta \hat{s}_R\p_+\hat{s}_R +i\delta \hat{s}_L\p_-\hat{s}_L \nn
& & 
+\p_\sigma\left\{-\frac{1}{2}\delta x^i\p_\sigma x^i
-\frac{1}{2}\delta x^a\p_\sigma x^a
+\frac{i}{2}(S_R\delta S_R-S_L\delta S_L)\right\}
\Bigg].
\eea
From this, equations of motion and boundary condition for open 
strings read as follows.

equations of motion
\bea
\p_+\p_-x^i+m^2x^i=0, \\
\p_+s_R-m\Lambda s_L=0, \\
\p_-s_L-m\Lambda s_R=0.
\eea
boundary conditions at $\sigma=0,\pi$
\bea
& & {\rm N} : \p_\sigma x^i=0, \\
& & {\rm D} : x^i={\rm const.}, \\
& & S_L=M S_R,\quad M^TM=1 \label{bcrrf}.
\eea
Note that in the light cone gauge $u$ and $v$ are always Neumann.

We consider only the case where both ends of open strings satisfy 
the same boundary condition.

Mode expansion of $x^i$ and its Hamiltonian are as follows.
\begin{itemize}
\item $x^i$ : N \\
\bea
x^i=i\sqrt{\frac{\ap}{2}}\left[
\sqrt{\frac{2}{\omega_0}}
(a^i_0e^{-i\omega_0\tau}-a_0^{i\dagger} e^{i\omega_0\tau})
+\sum_{n=1}^\infty\frac{1}{\sqrt{\omega_n}}
(a^i_ne^{-i\omega_n\tau}-a_n^{i\dagger} e^{i\omega_n\tau})
(e^{in\sigma}+e^{-in\sigma})\right],
\eea
\bea
& & \omega_n={\rm sgn}(n)\sqrt{n^2+(2m)^2}\quad (n\not\neq 0),
\quad \omega_0=2m, \\
& & \quad [a^i_n,a^{j\dagger}_{n'}]=\delta^{i,j}\delta_{n,n'},
\eea
\bea
H & = & \frac{1}{2\pi\ap}\int d\sigma\left[\Half \p_\tau x^i\p_\tau x^i
+\Half \p_\sigma x^i\p_\sigma x^i+2m^2x^ix^i\right] \nn
 & = & \Half\sum_{n=0}^\infty\omega_n(a_n^ia_n^{i\dagger}+a_n^{i\dagger}a_n^i).
\eea
\item $x^i$ : D \\
\bea
x^i & = & \frac{x_1^i(e^{2m\sigma}-e^{-2m\sigma})-
 x_0^i(e^{2m(\sigma-\pi)}-e^{-2m(\sigma-\pi)})}{e^{2m\pi}-e^{-2m\pi}}
\nn
 & & +i\sqrt{\frac{\ap}{2}}\sum_{n=1}^\infty\frac{1}{\sqrt{\omega_n}}
(a_n^ie^{-i\omega_n\tau}+a_n^{i\dagger} e^{i\omega_n\tau})
(e^{in\sigma}-e^{-in\sigma}),
\eea
\beq
[a^i_n,a^{j\dagger}_{n'}]=\delta^{i,j}\delta_{n,n'},
\eeq
where $x_0^i=x^i(\sigma=0)$, $x_1^i=x^i(\sigma=\pi)$.
\bea
H=\frac{m}{2\pi\ap}\frac{(e^{2m\pi}+e^{-2m\pi})
((x_1^i)^2+(x_0^i)^2)-4x_1^ix_0^i}{e^{2m\pi}-e^{-2m\pi}}
+\Half\sum_{n=1}^\infty\omega_n
(a^i_na_n^{i\dagger}+a_n^{i\dagger} a^i_n).
\eea
\end{itemize}

Next we give mode expansions and Hamiltonians of fermions. 
We assume that $M$ is given by a product of $\gamma^i$'s. Then 
the number of $\gamma^i$'s in $M$ must be even, since $S_L$ and $S_R$ have
the same chirality.
In addition since we assume $MR_\pm =R_\pm M$ later, 
we impose this condition from the beginning.
We define sign factors $\eta, \xi^i=\pm 1$ as follows,
\beq
\gamma^iM=\xi^iM\gamma^i, M^T=\eta M,
\eeq
and decompose $M$ into $4\times 4$ matrices:
\beq
M S_R=\left(\begin{array}{c}
Ns_{R,L} \\
\hat{N}\hat{s}_{R,L}
\end{array}\right).
\eeq
Then, noting that $(N\Lambda)^2=-\eta\xi_1\xi_2=\pm 1$,
\begin{itemize}
\item $\eta\xi_1\xi_2=+1$
\bea
s_L & = & \frac{\sqrt{\ap}}{2}\Bigg[\sqrt{2}
(\cos 2m\tau+\sin 2m\tau N\Lambda)s_0
\nn
 & & +\sum_{n=-\infty, n\not\neq 0}^\infty
e^{-i\omega_n\tau}
\left(\sqrt{\frac{\omega_n+n}{\omega_n}}e^{-in\sigma}
-{\rm sgn}(n)i\sqrt{\frac{\omega_n-n}{\omega_n}}
e^{in\sigma}N\Lambda\right)
s_n\Bigg],
\eea
\bea
s_R & = & \frac{\sqrt{\ap}}{2}\Bigg[\sqrt{2}
(\sin 2m\tau\Lambda+\eta\cos 2m\tau N)s_0
\nn
 & & +\sum_{n=-\infty, n\not\neq 0}^\infty
e^{-i\omega_n\tau}
\left(\eta\sqrt{\frac{\omega_n+n}{\omega_n}}e^{in\sigma}N
-{\rm sgn}(n)i\sqrt{\frac{\omega_n-n}{\omega_n}}
e^{-in\sigma}\Lambda\right)
s_n\Bigg],
\eea
\beq
\{s_n^\alpha,s_{n'}^\beta\}=\delta^{\alpha,\beta}\delta_{n,n'},\quad
s_n^\dagger=s_{-n},
\eeq
\bea
H & = & \frac{1}{2\pi\ap}\int_0^\pi d\sigma[is_L\dot{s}_L+is_R\dot{s}_R]
\nn
& = & -ims_0N\Lambda s_0+\Half\sum_{n=-\infty, n\not\neq 0}^\infty
 \omega_ns_{-n}s_n.
\eea

\item $\eta\xi_1\xi_2=-1$
\bea
s_L & = & \sqrt{\frac{\pi\ap m}{\sinh(2\pi m)}}
[\cosh(m(2\sigma-\pi))\eta N+\sinh(m(2\sigma-\pi))\Lambda]\Sigma \nn
& & +\frac{\sqrt{\ap}}{2}\sum_{n=-\infty, n\not\neq 0}^\infty
e^{-i\omega_n\tau}
\Bigg(\sqrt{\frac{\omega_n+n}{\omega_n}}e^{-in\sigma} \nn
& & 
-{\rm sgn}(n)\frac{1}{\omega_n}\sqrt{\frac{\omega_n-n}{\omega_n}}
(2m+inN\Lambda)e^{in\sigma}\Bigg)s_n,
\eea
\bea
s_R & = & \sqrt{\frac{\pi\ap m}{\sinh(2\pi m)}}
[\cosh(m(2\sigma-\pi))+\sinh(m(2\sigma-\pi))N\Lambda]\Sigma \nn
& & +\frac{\sqrt{\ap}}{2}\sum_{n=-\infty, n\not\neq 0}^\infty
e^{-i\omega_n\tau}
\Bigg({\rm sgn}(n)i\sqrt{\frac{\omega_n-n}{\omega_n}}e^{-in\sigma}\Lambda
\nn
& & 
+\frac{1}{\omega_n}(-2mi\Lambda+\eta nN)\sqrt{\frac{\omega_n+n}{\omega_n}}
e^{in\sigma}\Bigg)s_n,
\eea
\beq
\{s_n^\alpha,s_{n'}^\beta\}=\delta^{\alpha,\beta}\delta_{n,n'},\quad
s_n^\dagger=s_{-n},\quad
\{\Sigma^\alpha,\Sigma^\beta\}=\delta^{\alpha,\beta},\quad
\Sigma^\dagger=\Sigma,
\eeq
\beq
H=\Half\sum_{n=-\infty, n\not\neq 0}^\infty\omega_ns_{-n}s_n.
\eeq
\end{itemize}

Now we investigate the condition that half of the dynamical supersymmetry
are unbroken.
From eq.(\ref{bcrrf}) the following condition must be satisfied. 
\beq
\delta S_L=M \delta S_R \quad{\rm at }\quad\sigma=0,\pi.
\eeq
We require the first, second and third terms of (\ref{dsrrsr}) 
and (\ref{dsrrsl}) cancel separately, for each $i$ and $a$. 
Then $M$ must satisfy $MR_\pm =R_\pm M$, and
\beq
M\delta S_R = -\eta\xi^i\p_-x^i(\epsilon_R M\gamma^iR_+)
-\eta\xi^a\p_-x^a(\epsilon_R M\gamma^aR_-)
-\eta\xi^i\xi^1\xi^2 mx^i(\epsilon_L M\gamma^i\gamma^1\gamma^2R_+).
\label{mdsrrdr}
\eeq
Comparing the above with (\ref{dsrrsl}), unbroken supersymmetry 
parameters are given in the following form.
\beq
\epsilon_R M=\zeta\epsilon_L,\quad\zeta=\pm 1.
\eeq

Equating the first terms of (\ref{mdsrrdr}) and (\ref{dsrrsl}),
we find the following result.
\begin{itemize}
\item $\zeta\eta\xi^i=+1$ \\
 $x^i$ must be Neumann.
 Then the third terms of (\ref{mdsrrdr}) and (\ref{dsrrsl}) give 
 an additional constraint: 
 $\zeta\xi^1\xi^2\xi^i=+1$ i.e. $\eta\xi^1\xi^2=+1$, otherwise $x^i=0$, 
 which contradicts the Neumann condition.
\item $\zeta\eta\xi^i=-1$ \\
 $x^i$ must be Dirichlet. Then by the cancellation of the third terms,
 \begin{enumerate}
 \item $\zeta\xi^1\xi^2\xi^i=+1$ i.e. $\eta\xi^1\xi^2=-1$ \\
  Positions of D-branes can be arbitrary.
 \item $\zeta\xi^1\xi^2\xi^i=-1$ i.e. $\eta\xi^1\xi^2=+1$ \\
  D-branes must be at $x^i=0$
 \end{enumerate}
\end{itemize}
Analysis for $x^a$ is similar, but without additional constraint coming from
the third terms of (\ref{dsrrsr}) and (\ref{dsrrsl}).
It is shown easily that the supersymmetry transformation of $x^i$ vanish 
at the boundary if $x^i$ is Dirichlet.

We can classify $M$ and the result, up to symmetry rotation and 
$\gamma^1\dots\gamma^8=1$, is given 
as follows.
\begin{itemize}
\item $M=1$ ($\eta=1, \xi_1\xi_2=1$)\\
\begin{tabular}{lll}
 $\zeta=+1$ & $x^{1,2,3,4,5,6,7,8}$ : N & (D9-brane) \\
 $\zeta=-1$ & $x^{1,2,3,4,5,6,7,8}$ : D ($x^i=0$) & (D1-brane)
\end{tabular}
\item $M=\gamma^{13}$ ($\eta=-1, \xi_1\xi_2=-1$)\\
\begin{tabular}{lllll}
 $\zeta=+1$ & $x^{1,3}$ : N, & $x^{2,4}$ : D ($x^i=0$), & $x^{5,6,7,8}$ : D
 & (D3-brane)\\
 $\zeta=-1$ & $x^{1,3}$ : D ($x^i=0$), & $x^{2,4}$ : N, & $x^{5,6,7,8}$ : N
 & (D7-brane)
\end{tabular}
\item $M=\gamma^{14}$ ($\eta=-1, \xi_1\xi_2=-1$)\\
\begin{tabular}{lllll}
 $\zeta=+1$ & $x^{1,4}$ : N, & $x^{2,3}$ : D ($x^i=0$), & $x^{5,6,7,8}$ : D
 & (D3-brane)\\
 $\zeta=-1$ & $x^{1,4}$ : D ($x^i=0$), & $x^{2,3}$ : N, & $x^{5,6,7,8}$ : N
 & (D7-brane)
\end{tabular}
\item $M=\gamma^{56}$ ($\eta=-1, \xi_1\xi_2=1$)\\
\begin{tabular}{lllll}
 $\zeta=+1$ & $ x^{1,2,3,4}$ : D, & $x^{5,6}$: N, & $x^{7,8}$: D & (D3-brane) 
\end{tabular}
\item $M=\gamma^{1234}$ ($\eta=1, \xi_1\xi_2=1$)\\
\begin{tabular}{llll}
 $\zeta=+1$ & $x^{1,2,3,4}$ : D ($x^i=0$), & $x^{5,6,7,8}$ : N & (D5-brane)\\
 $\zeta=-1$ & $x^{1,2,3,4}$ : N, & $x^{5,6,7,8}$ : D & (D5-brane)
\end{tabular}
\item $M=\gamma^{1256}$ ($\eta=1, \xi_1\xi_2=1$)\\
\begin{tabular}{llllll}
 $\zeta=+1$ & $x^{1,2}$ : D ($x^i=0$), & $x^{3,4}$ : N, & $x^{5,6}$ : D, 
 & $x^{7,8}$ : N & (D5-brane)\\
 $\zeta=-1$ & $x^{1,2}$ : N, & $x^{3,4}$ : D ($x^i=0$), & $x^{5,6}$ : N,
 & $x^{7,8}$ : D & (D5-brane)
\end{tabular}
\end{itemize}

Note that except $M=\gamma^{56}$ case D-branes sit at $x^i=0$.
This phenomenon is similar to the result for pp-wave obtained from 
$AdS_5\times S^5$ \cite{bp,dp}. $M=\gamma^{1234}$ case has been already 
analyzed in \cite{kns}. 
We note for future reference that $M=\gamma^{12}$ also gives 
D3-branes, though this case is not allowed by the third terms of 
(\ref{dsrrsr}) and (\ref{dsrrsl}).

Next let us consider the condition that some of kinematical 
supersymmetry are unbroken. It is given by
$\tilde{\delta}S_L=M\tilde{\delta}S_R$.
We decompose $\tilde{\epsilon}_{R,L}$ into 4-component spinors 
in the same way as (\ref{decomp}):
\bea
\tilde{\epsilon}_{R,L}=\left(\begin{array}{c} \kappa_{R,L} \\ 
\hat{\kappa}_{R,L} \end{array}\right).
\eea
Then,
\bea
& & (1-N)e^{2m\tau\Lambda}\kappa_R=(1+N)e^{-2m\tau\Lambda}\kappa_L, \\
& & (1-\hat{N})\hat{\kappa}_R=(1+\hat{N})\hat{\kappa}_L.
\eea
Noting that $\hat{N}^2=\pm 1$ the second equation means half of 
$\hat{\kappa}_R$ and $\hat{\kappa}_L$ are unbroken.
For the first equation we consider the following two cases separately.
\begin{itemize}
\item $N^2(=\eta)=1$ \\
The left and right hand side must vanish separately:
\beq
(1-N)e^{2m\tau\Lambda}\kappa_R=(1+N)e^{-2m\tau\Lambda}\kappa_L=0.
\eeq
These equations are satisfied for arbitrary $\tau$ only when
$N\Lambda=\Lambda N$.
Then half of $\kappa_R$ and $\kappa_L$ are unbroken.
\item $N^2=-1$ \\
The following must be satisfied.
\beq
e^{2m\tau\Lambda}\kappa_R=Ne^{-2m\tau\Lambda}\kappa_L.
\eeq
If $N\Lambda=-\Lambda N$ this is satisfied for arbitrary $\tau$ and
half of $\kappa_R$ and $\kappa_L$ are unbroken. If $N\Lambda=\Lambda N$
this is not satisfied.
\end{itemize}
The above classification is summarized as follows.
If $(N\Lambda)^2 (=-\eta\xi_1\xi_2)=-1$, 
half of $\kappa_R$ and $\kappa_L$ are unbroken.
If $(N\Lambda)^2=1$, none of $\kappa_R$ and $\kappa_L$ are unbroken.
By this analysis we find that $M=\gamma^{56}$ (and $\gamma^{12}$)
preserves 1/4 of kinematical supersymmetry, and $M=1, \gamma^{13}, 
\gamma^{14}, \gamma^{1234}, \gamma^{1256}$ preserve half.

\section{D-branes in NSNS pp-wave background}

In this section we consider D-branes on the NSNS pp-wave background
which is S-dual to the RR background in the previous section\cite{rt}:
\bea
ds^2=dudv-f^2x^ix^idu^2+dx^idx^i+dx^adx^a, \nn
B_{ui_1}=-\Half{\cal F}_{i_1j_1}x^{j_1},\quad
B_{ui_2}=-\Half{\cal F}_{i_2j_2}x^{j_2},
\eea
where $B$ is the NSNS 2-form. We will use the same notation as in the 
previous section.

Worldsheet action of fundamental strings in GS formalism is
\bea
L & = & \frac{1}{\pi\ap}\Bigg[\Half\p_+u\p_-v+\Half\p_+v\p_-u-m^2x^ix^i \nn
& & +\Half{\cal F}_{ij}x^i(\p_+u\p_-x^j-\p_-u\p_+x^j)
+\p_+x^i\p_-x^i+\p_+x^a\p_-x^a \nn
& & +iS_R(\p_+-\frac{1}{8}\ap p^u{\cal F}_{ij}\gamma^{ij})S_R
+iS_L(\p_-+\frac{1}{8}\ap p^u{\cal F}_{ij}\gamma^{ij})S_L
\Bigg] \nn
& = & \frac{1}{\pi\ap} [ \p_+(e^{2im\sigma}X)\p_-(e^{-2im\sigma}X^\dagger)
+\p_+(e^{-2im\sigma}X^\dagger)\p_-(e^{2im\sigma}X) \nn
& & +iS_Re^{2m\tau\gamma^1\gamma^2 R_+}\p_+
(e^{-2m\tau\gamma^1\gamma^2 R_+}S_R)
+iS_Le^{-2m\tau\gamma^1\gamma^2 R_+}\p_-(e^{2m\tau\gamma^1\gamma^2 R_+}S_L)
],
\eea
where 
$X=\frac{1}{\sqrt{2}}(x^1+ix^2)$ 
and other bosonic fields are ignored in 
the second expression.
Though this system can also be analyzed by covariant NSR 
formalism\cite{tt,hs},
we adopt GS formalism in order to compare with the RR background.

Two types of supersymmetry are as follows.
\begin{itemize}
\item dynamical supersymmetry
\bea
\delta x^i & = & i\epsilon_R\gamma^iR_+S_R+i\epsilon_L\gamma^iR_+S_L, \nn
\delta x^a & = & i\epsilon_R\gamma^iR_-S_R+i\epsilon_L\gamma^iR_-S_L, \nn
\delta S_R & = & -\p_-x^i\epsilon_R\gamma^iR_+-\p_-x^a\epsilon_R\gamma^aR_-
 +mx^i\epsilon_R\gamma^i\gamma^1\gamma^2R_+, \nn
\delta S_L & = & -\p_+x^i\epsilon_L\gamma^iR_+-\p_+x^a\epsilon_L\gamma^aR_-
 -mx^i\epsilon_L\gamma^i\gamma^1\gamma^2R_+,
\eea
where $\epsilon_{R,L}R_+=0$.
\item kinematical supersymmetry
\bea
\delta S_R & = & e^{2m\tau\gamma^1\gamma^2 R_+}\epsilon_R, \nn
\delta S_L & = & e^{-2m\tau\gamma^1\gamma^2 R_+}\epsilon_L.
\eea
\end{itemize}

As in the previous section, we do not add extra boundary term 
in this section. Then variation of the action is,
\bea
\delta S & = & \frac{1}{\pi\ap}\int d^2\sigma\Bigg[
-2\delta x^{i_1}(\p_+\p_-x^{i_1}+m^2x^{i_1}
-m\epsilon_{i_1j_1}(\p_-x^{j_1}-\p_+x^{j_1})) \nn
& & -2\delta x^{i_2}(\p_+\p_-x^{i_2}+m^2x^{i_2}
-m\epsilon_{i_2j_2}(\p_-x^{j_2}-\p_+x^{j_2})) \nn
& & +2i\delta s_Re^{-2m\tau\Lambda}\p_+
(e^{2m\tau\Lambda}s_R)
+2i\delta s_Le^{2m\tau\Lambda}\p_-
(e^{-2m\tau\Lambda}s_L) \nn
& & +\p_\sigma\left\{-\Half\delta x^{i_1}
(\p_\sigma x^{i_1}-2m\epsilon_{i_1j_1}x^{j_1})
-\Half\delta x^{i_2}
(\p_\sigma x^{i_2}-2m\epsilon_{i_2j_2}x^{j_2})
+\frac{i}{2}(S_R\delta S_R-S_L\delta S_L)\right\}
\Bigg].
\eea

Equations of motion are
\bea
& & \p_+\p_-x^{i_1}+m^2x^{i_1}-m\epsilon^{i_1j_1}
 (\p_-x^{j_1}-\p_+x^{j_1})=0, \\
& & \p_+(e^{2m\tau\Lambda}s_R)=0, \\
& & \p_-(e^{-2m\tau\Lambda}s_L)=0.
\eea

Boundary conditions for open strings are
\bea
& & {\rm N} : \p_\sigma x^{i_1}-2m\epsilon^{i_1j_1}x^{j_1}=0, 
\label{nsneumann}\\
& & {\rm D} : x^i={\rm const.}, \\
& & S_L=M S_R,\quad M^TM=1.
\eea

In terms of $X$ the equation of motion and the boundary condition are
\bea
& & \p_+\p_-(e^{2im\sigma}X)=0, \\
& & x^1,x^2 : {\rm N}\quad \p_\sigma(e^{2im\sigma}X)=0, \\
& & x^1,x^2 : {\rm D}\quad e^{2im\sigma}X={\rm const.}.
\eea
Thus if both $x^1$ and $x^2$ are Neumann or Dirichlet, then $e^{2im\sigma}X$
satisfy equation of motion and Neumann or Dirichlet
condition in the flat background. Therefore the analysis is almost the same
as in the flat background.

Mode expansion and Hamiltonian of $x^i$ is given as follows.
\begin{itemize}
\item $x^1,x^2$ : N
\beq
X=e^{-2im\sigma}\left[x+2\ap p\tau+
 i\sqrt{\frac{\ap}{2}}\sum_{n\not\neq 0}\frac{1}{n}\alpha_n
 (e^{-in(\tau-\sigma)}+e^{-in(\tau+\sigma)})\right],
\eeq
\beq
[x,p^\dagger]=i,\quad [\alpha_n,\alpha_{n'}^\dagger]=n\delta_{nn'},
\eeq
\bea
H & = & \frac{1}{\pi\ap}\int_0^\pi d\sigma\left[\p_+(e^{2im\sigma}X)\p_+
(e^{-2im\sigma}X^\dagger)
 +\p_-(e^{2im\sigma}X)\p_-(e^{-2im\sigma}X^\dagger)\right] \nn
& = & 2\ap pp^\dagger+\Half\sum_{n\not\neq 0}(\alpha_n\alpha_n^\dagger
+\alpha_n^\dagger\alpha_n).
\eea
\item $x^1,x^2$ : D
\beq
X=e^{-2im\sigma}\Bigg[x_0+(x_1e^{2im\pi}-x_0)\frac{\sigma}{\pi}
 +i\sqrt{\frac{\ap}{2}}\sum_{n\not\neq 0}\frac{1}{n}\alpha_n
 (e^{-in(\tau-\sigma)}-e^{-in(\tau+\sigma)})\Bigg],
\eeq
\beq
[\alpha_n,\alpha_{n'}^\dagger]=n\delta_{nn'},
\eeq
where $x_0=X(\sigma=0), x_1=X(\sigma=\pi)$.
\beq
H=\frac{1}{2\pi^2\ap}(x_1e^{2im\pi}-x_0)(x_1e^{2im\pi}-x_0)^*
+\Half\sum_{n\not\neq 0}(\alpha_n\alpha_n^\dagger+\alpha_n^\dagger\alpha_n).
\eeq
\item $x^1$ : N, $x^2$ : D
\bea
X & = & e^{-2im\sigma}\Bigg[\sqrt{2}i
\frac{x_0e^{2\pi im}-x_1}{e^{2\pi im}-e^{-2\pi im}} \nn
& & +i\sqrt{\frac{\ap}{2}}\sum_{n=-\infty}^\infty 
\left(\frac{1}{n+2m}\alpha_n e^{-i(n+2m)(\tau-\sigma)}
+\frac{1}{n-2m}\alpha_{-n}^\dagger e^{-i(n-2m)(\tau+\sigma)}
\right)\Bigg],
\eea
where $x_0=x^2(\sigma=0), x_1=x^2(\sigma=\pi)$.
(Here we assume $2m$ is not an integer. 
If $2m$ is integer, $X$ has extra zero mode contribution.)
\beq
[\alpha_n,\alpha_{n'}^\dagger]=(n+2m)\delta_{nn'},
\eeq
\beq
H=\Half\sum_{n}(\alpha_n\alpha_n^\dagger+\alpha_n^\dagger\alpha_n).
\eeq
\end{itemize}

Next we give mode expansions and Hamiltonians of fermions.
The boundary condition is
\beq
s_L=Ns_R\quad {\rm at}\quad\sigma=0,\pi. 
\eeq
Note that $N\Lambda=\xi_1\xi_2\Lambda N$. Then,
\begin{itemize}
\item $\xi_1\xi_2=-1$
\bea
s_L=\sqrt{\frac{\ap}{2}}e^{2m\tau\Lambda}N
\sum_{n=-\infty}^\infty s_ne^{-in(\tau+\sigma)}, \nn
s_R=\sqrt{\frac{\ap}{2}}e^{-2m\tau\Lambda}
\sum_{n=-\infty}^\infty s_ne^{-in(\tau-\sigma)}, 
\eea
\beq
\{s_n^\alpha,s_{n'}^{\beta\dagger}\}=\delta_{nn'}\delta^{\alpha\beta},
s_n^\dagger=s_{-n},
\eeq
\bea
H & = & \frac{i}{2\pi\ap}\int_0^\pi d\sigma(s_R\dot{s}_R+s_L\dot{s}_L) \nn
& = & \Half\sum_{n=-\infty}^\infty s_{-n}(n-2im\Lambda)s_n.
\eea
\item $\xi_1\xi_2=1$
\bea
s_L=\sqrt{\frac{\ap}{2}}Ne^{-2m\sigma\Lambda}
\sum_{n=-\infty}^\infty s_ne^{-in(\tau+\sigma)}, \nn
s_R=\sqrt{\frac{\ap}{2}}e^{-2m\sigma\Lambda}
\sum_{n=-\infty}^\infty s_ne^{-in(\tau-\sigma)}, 
\eea
\beq
\{s_n^\alpha,s_{n'}^{\beta\dagger}\}=\delta_{nn'}\delta^{\alpha\beta},
s_n^\dagger=s_{-n},
\eeq
\beq
H=\Half\sum_{n=-\infty}^\infty ns_{-n}s_n.
\eeq
\end{itemize}

Now we consider the half dynamical supersymmetry condition.
Supersymmetry transformation of $S_R$ and $S_L$ can be rewritten as follows.
\bea
\delta S_R & = & 
-\Half(\p_\tau x^i-(\p_\sigma x^i-2m\epsilon_{i_1j_1} x^{j_1}))
\epsilon_R\gamma^iR_+
-\p_-x^a\epsilon_R\gamma^aR_-, \nn
\delta S_L & = & 
-\Half(\p_\tau x^i+(\p_\sigma x^i-2m\epsilon_{i_1j_1} x^{j_1}))
\epsilon_L\gamma^iR_+
-\p_+x^a\epsilon_L\gamma^aR_-.
\eea
Noting that Neumann condition is given by eq.(\ref{nsneumann}), 
we can see from the above form of supersymmetry transformation that
BPS condition is the same as $m=0$ case i.e. flat background,
except that we have an additional condition $MR_\pm=R_\pm M$.

For example, $M=\gamma^{i_1}\dots\gamma^{i_{p-1}}$ gives Dp-brane 
lying along $x^{i_1}\dots x^{i_{p-1}}$.
Note that there is no restriction for positions of D-branes.

Next let us consider half kinematical supersymmetry condition.
\bea
& & e^{2m\tau\Lambda}\kappa_L=Ne^{-2m\tau\Lambda}\kappa_R, \\
& & \hat{\kappa}_L=\hat{N}\hat{\kappa}_R.
\eea
The first equation can be satisfied for arbitrary $\tau$ only when 
$\Lambda N=-N\Lambda$ i.e. $\xi_1\xi_2=-1$.
Then the amount of unbroken kinematical supersymmetry is half, otherwise 1/4.

\section{D-branes in RR background with gauge flux and S-duality}

In the previous two sections we have investigated boundary conditions
in the RR and NSNS backgrounds respectively, without introducing 
additional boundary term. These backgrounds are 
S-dual to each other, and D3-branes in one of these background are also 
S-dual to D3-branes in 
the other background. However, we have seen that D3-branes in the RR 
background can sit only at $x^i=0$ (except the case $M=\gamma^{56}$), 
and in the NSNS background they can be anywhere. 
Note that the NSNS background has nonzero NSNS B-field, which induces 
gauge flux on D-branes. Therefore D3-branes in section 3 are S-dual to 
D3-branes with nonzero gauge flux in the RR background.
This means that if we turn on appropriate gauge field on 
D-branes in the RR background, they can move away from $x^i=0$ without
breaking any more supersymmetry. The purpose of this section is to 
construct such D3-branes in the RR background.
Such a phenomenon has been already pointed out in \cite{st} for pp-wave 
from $AdS_5\times S^5$.

What gauge field should be turned on can be seen 
by noticing that S-duality is realized as standard
electromagnetic duality on D3-brane effective action.
The effective action in nonzero NSNS and RR 2-form background is 
\bea
S & = & -\int d^4\sigma\Bigg[\sqrt{-{\rm det}
(\p_\mu X^M\p_\nu X^N(G_{MN}+B_{MN})+F_{\mu\nu})} \nn
& & 
\pm\frac{1}{4}\epsilon^{\mu\nu\lambda\rho}C_{MN}\p_\mu X^M\p_\nu X^N
(F_{\lambda\rho}+B_{KL}\p_\lambda X^K\p_\rho X^L)\Bigg].
\eea
To perform dual transformation, we add $\pm\int d^4\sigma\Half
\epsilon^{\mu\nu\lambda\rho}\p_\mu\lambda_\nu F_{\lambda\rho}$ 
to the above action, where $\lambda$ is a Lagrange multiplier, 
which will be the gauge field of the dualized system.
Then equation of motion obtained by varying $F_{\mu\nu}$ is
\bea
\p_\mu\lambda_\nu-\p_\nu\lambda_\mu-C_{MN}\p_\mu X^M\p_\nu X^N
 & = & \pm\frac{1}{4}\epsilon_{\mu\nu\lambda\rho}\sqrt{-{\rm det}
(\p_\mu X^M\p_\nu X^N(G_{MN}+B_{MN})+F_{\mu\nu})} \nn
& & \times 
[(\p_\mu X^M\p_\nu X^N(G_{MN}+B_{MN})+F_{\mu\nu})^{-1 \lambda\rho} \nn
& & -(\p_\mu X^M\p_\nu X^N(G_{MN}+B_{MN})+F_{\mu\nu})^{-1 \rho\lambda}].
\eea
The left hand side of this equation is ``$F+B$'' of the dualized system.
Therefore we can read off what gauge field must be turned on by computing 
the right hand side. We give results for the case where D3-branes
lie along $(u,v,x^1,x^2)$, $(u,v,x^1,x^3)$ and $(u,v,x^1,x^4)$. 
\begin{itemize}
\item $(u,v,x^1,x^2) (M=\gamma^{12})$ \\
\beq
F_{ui}=\pm fx^i.
\eeq
Neumann boundary condition : 
$\p_\sigma x^i-F_{iu}\p_\tau u=\p_\sigma x^i\pm 2mx^i=0.$
\item $(u,v,x^1,x^3) (M=\gamma^{13})$ \\
\beq
F_{u1}=\mp fx^4,\quad F_{u3}=\pm fx^2.
\eeq
Neumann boundary condition : 
\bea
& & \p_\sigma x^1-F_{1u}\p_\tau u=\p_\sigma x^1\mp 2mx^4=0, \nn
& & \p_\sigma x^3-F_{3u}\p_\tau u=\p_\sigma x^3\pm 2mx^2=0. \nonumber
\eea
\item $(u,v,x^1,x^4) (M=\gamma^{14})$ \\
\beq
F_{u1}=\pm fx^3,\quad F_{u4}=\pm fx^2
\eeq
Neumann boundary condition : 
\bea
& & \p_\sigma x^1-F_{1u}\p_\tau u=\p_\sigma x^1\pm 2mx^3=0, \nn
& & \p_\sigma x^4-F_{4u}\p_\tau u=\p_\sigma x^4\pm 2mx^2=0. \nonumber
\eea
\end{itemize}

We can also see these gauge fluxes directly from the half dynamical 
supersymmetry conditions. First let us consider the case $M=\gamma^{13}$.
By using $\epsilon_RM=\zeta\epsilon_L$, the supersymmetry transformation of 
$S_R$ and $S_L$ can be rewritten as follows.
\bea
M\delta S_R & = & \zeta(\p_-x^1+\zeta mx^4)\epsilon_L\gamma^1R_+
-\zeta(\p_-x^2-\zeta mx^3)\epsilon_L\gamma^2R_+ \nn
& & +\zeta(\p_-x^3-\zeta mx^2)\epsilon_L\gamma^3R_+
-\zeta(\p_-x^4+\zeta mx^1)\epsilon_L\gamma^4R_+,
\eea
\bea
\delta S_L & = & -(\p_+x^1-\zeta mx^4)\epsilon_L\gamma^1R_+
-(\p_+x^2+\zeta mx^3)\epsilon_L\gamma^2R_+ \nn
& & -(\p_+x^3+\zeta mx^2)\epsilon_L\gamma^3R_+
-(\p_+x^4-\zeta mx^1)\epsilon_L\gamma^4R_+.
\eea
Therefore boundary conditions are
\begin{itemize}
\item $\zeta=-1$ \\
$\p_\sigma x^1+2mx^4=0,\quad\p_\sigma x^3-2mx^2=0,\quad
\p_\tau x^2=0,\quad\p_\tau x^4=0$.
\item $\zeta=+1$ \\
$\p_\sigma x^2+2mx^3=0,\quad\p_\sigma x^4-2mx^1=0,\quad
\p_\tau x^1=0,\quad\p_\tau x^3=0$.
\end{itemize}
Note that Dirichlet condition has no additional constraint such as $x^i=0$.
Therefore D-branes can be put anywhere.

We give mode expansions and Hamiltonians for the following example

\begin{tabular}{ll}
N : & $\p_\sigma x^1+2mx^4=0$ \\
D : & $\p_\tau x^4=0$
\end{tabular}

\bea
x^1 & = & -\frac{x_1-x_0e^{-2m\pi}}{e^{2m\pi}-e^{-2m\pi}}e^{2m\sigma}
+\frac{x_0e^{2m\pi}-x_1}{e^{2m\pi}-e^{-2m\pi}}e^{-2m\sigma} \nn
& & +i\sqrt{\ap}\frac{1}{\sqrt{\omega_0}}
(a^1_0e^{-i\omega_0\tau}-a_0^{1\dagger} e^{i\omega_0\tau}) \nn
& & +i\sqrt{\frac{\ap}{2}}\sum_{n=1}^\infty\frac{1}{\sqrt{\omega_n}}
(a^1_ne^{-i\omega_n\tau}-a_n^{1\dagger} e^{i\omega_n\tau})
(e^{in\sigma}+e^{-in\sigma}),
\eea
\bea
x^4 & = & \frac{x_1-x_0e^{-2m\pi}}{e^{2m\pi}-e^{-2m\pi}}e^{2m\sigma}
+\frac{x_0e^{2m\pi}-x_1}{e^{2m\pi}-e^{-2m\pi}}e^{-2m\sigma} \nn
& & +i\sqrt{\frac{\ap}{2}}\sum_{n=1}^\infty\frac{1}{\sqrt{\omega_n}}
(a^4_ne^{-i\omega_n\tau}+a_n^{4\dagger} e^{i\omega_n\tau})
(e^{in\sigma}-e^{-in\sigma}),
\eea
\beq
[a_n^i,a_{n'}^{j\dagger}]=\delta_{nn'}\delta^{ij},
\eeq
where $x_0=x^4(\sigma=0), x_1=x^4(\sigma=\pi)$.

Hamiltonian has an additional factor 
$\frac{1}{2\pi\ap}[-2\zeta mx^1x^4]_{\sigma=0}^{\sigma=\pi}$ 
originated from the gauge field.
\bea
H & = & \frac{1}{2\pi\ap}\int_0^\pi d\sigma\left[\Half\p_\tau x^1\p_\tau x^1 
+\Half\p_\tau x^4\p_\tau x^4 
+\Half\p_\sigma x^1\p_\sigma x^1 
+\Half\p_\sigma x^4\p_\sigma x^4 
+2m^2x^1x^1+2m^2x^4x^4\right] \nn
& & +\frac{1}{2\pi\ap}[-2\zeta mx^1x^4]_{\sigma=0}^{\sigma=\pi}
\nn
& = & \frac{1}{2\pi\ap}\int_0^\pi d\sigma\left[\Half\p_\tau x^1\p_\tau x^1 
+\Half\p_\tau x^4\p_\tau x^4 
+\Half(\p_\sigma x^1-2\zeta mx^4)^2
+\Half(\p_\sigma x^4-2\zeta mx^1)^2\right] \nn
& = & \Half\left[\omega_0(a^1_0a_0^{1\dagger}+a_0^{1\dagger} a^1_0)
+\sum_{n=1}^\infty\omega_n(a^1_na_n^{1\dagger}+a_n^{1\dagger}a^1_n
+a^4_na_n^{4\dagger}+a_n^{4\dagger}a^4_n)\right].
\eea

The fermion part is not changed since its boundary condition does not change.
Therefore the number of unbroken kinematical supersymmetry is the same as 
the case without gauge flux. Hence this D-brane is a half BPS state.

Next we investigate the case $M=\gamma^{14}$.
By using $\epsilon_RM=\zeta\epsilon_L$,
\bea
M\delta S_R & = & \zeta(\p_-x^1-\zeta mx^3)\epsilon_L\gamma^1R_+
-\zeta(\p_-x^2-\zeta mx^4)\epsilon_L\gamma^2R_+ \nn
& & -\zeta(\p_-x^3-\zeta mx^1)\epsilon_L\gamma^3R_+
+\zeta(\p_-x^4-\zeta mx^2)\epsilon_L\gamma^4R_+,
\eea
\bea
\delta S_L & = & -(\p_+x^1+\zeta mx^3)\epsilon_L\gamma^1R_+
-(\p_+x^2+\zeta mx^4)\epsilon_L\gamma^2R_+ \nn
& & -(\p_+x^3+\zeta mx^1)\epsilon_L\gamma^3R_+
-(\p_+x^4+\zeta mx^2)\epsilon_L\gamma^4R_+.
\eea
Boundary conditions are
\begin{itemize}
\item $\zeta=-1$ \\
$\p_\sigma x^2+2mx^4=0,\quad\p_\sigma x^3+2mx^1=0,\quad
\p_\tau x^1=0,\quad\p_\tau x^4=0$.
\item $\zeta=+1$ \\
$\p_\sigma x^1-2mx^3=0,\quad\p_\sigma x^4-2mx^2=0,\quad
\p_\tau x^2=0,\quad\p_\tau x^3=0$.
\end{itemize}
Mode expansions and Hamiltonians are similar to the previous case.

Finally let us consider the case $M=\gamma^{12}$, 
which is not allowed in the analysis in section 2.
By $\epsilon_RM=\zeta\epsilon_L$,
\beq
M\delta S_R = \zeta(\p_-x^{i_1}-\zeta mx^{i_1})
\epsilon_L\gamma^{i_1}R_+
-\zeta(\p_-x^{i_2}+\zeta mx^{i_2})\epsilon_L\gamma^{i_2}R_+,
\eeq
\beq
\delta S_L = -(\p_+x^{i_1}+\zeta mx^{i_1})
\epsilon_L\gamma^{i_1}R_+
-(\p_+x^{i_2}-\zeta mx^{i_2})\epsilon_L\gamma^{i_2}R_+.
\eeq
Boundary conditions are
\begin{itemize}
\item $\zeta=-1$ \\
 $\p_\sigma x^{i_1}-2mx^{i_1}=0$, $\p_\tau x^{i_2}=0$.
\item $\zeta=+1$ \\
 $\p_\tau x^{i_1}=0$, $\p_\sigma x^{i_2}-2mx^{i_2}=0$.
\end{itemize}
Thus this case is allowed and positions of D-branes can be arbitrary.

Mode expansion of $x$ which satisfy the boundary condition 
$\p_\sigma x-2mx=0$ is
\bea
x & = & \sqrt{\frac{4m\pi}{e^{4m\pi}-1}}
(x_0+2\ap p\tau)e^{2m\sigma} \nn
& & +i\sqrt{\frac{\ap}{2}}\sum_{n=1}^\infty
\frac{1}{\sqrt{\omega_n}}\Bigg[e^{-i\omega_n\tau}
\left(\frac{n-2mi}{\omega_n}e^{in\sigma}
+\frac{n+2mi}{\omega_n}e^{-in\sigma}\right)a_n \nn
& &
-e^{i\omega_n\tau}
\left(\frac{n-2mi}{\omega_n}e^{in\sigma}
+\frac{n+2mi}{\omega_n}e^{-in\sigma}\right)
a_n^\dagger
\Bigg],
\eea
\beq
[a_n,a_n^\dagger]=\delta_{n,n'},\quad [x_0,p]=i.
\eeq
Hamiltonian has an additional factor 
$\frac{1}{2\pi\ap}[-mxx]_{\sigma=0}^{\sigma=\pi}$ 
from the gauge field.
\bea
H & = & \frac{1}{2\pi\ap}\int_0^\pi d\sigma\left[\Half\p_\tau x\p_\tau x
+\Half\p_\sigma x\p_\sigma x +2m^2xx\right]
+\frac{1}{2\pi\ap}[-mxx]_{\sigma=0}^{\sigma=\pi}
\nn
& = & \frac{1}{2\pi\ap}\int_0^\pi d\sigma\left[\Half\p_\tau x\p_\tau x
+\Half(\p_\sigma x-2mx)^2\right] \nn
& = & \ap p^2+\Half\sum_{n=1}^\infty\omega_n(a_na_n^\dagger+a_n^\dagger a_n).
\eea

Thus we have constructed D3-branes corresponding to 
D3-branes in section 3 ($M=\gamma^{12}, \gamma^{13}, 
\gamma^{14}, \gamma^{56}$). For $M=\gamma^{56}$ we do not have to 
introduce nonzero gauge field. $M=\gamma^{13}$ and $\gamma^{14}$
give half BPS branes. $M=\gamma^{12}$ and $\gamma^{56}$ give 1/3 BPS 
branes (half dynamical and 1/4 kinematical supersymmetry).
It is expected that low lying open string spectra on these branes 
coincide with those of section 3 in $\ap\rightarrow 0$ limit.
Indeed we can see from the form of the Hamiltonians that 
the spectra of open strings with both ends ending on the same brane
agree.

\section{Conclusions}

We have investigated open string boundary conditions which preserve 
half dynamical supersymmetry and some of kinematical supersymmetry. 
Section 2 is devoted to D-branes without extra boundary term in the RR
background. These branes cannot move from $x^i=0$ in most cases.
Section 3 is devoted to D-branes without extra boundary term in the NSNS 
background. These branes can move away from $x^i=0$.
Section 4 is devoted to D-branes with gauge flux in the RR background, 
which correspond to those of section 2. These branes can indeed move 
away from $x^i=0$ with breaking no more supersymmetry. 

It is known that D-branes in pp-wave background from $AdS_5\times S^5$ 
can also move away from the origin by introducing gauge field\cite{st}. 
Similar phenomenon might happen in other pp-wave backgrounds.
It is interesting to investigate it by the same method as this paper or
Born Infeld action as in \cite{st}.

In \cite{bgg} it is shown that in the case of pp-wave from $AdS^5\times S^5$ 
only half BPS D-branes satisfy the Cardy condition 
and therefore less supersymmetric D-branes are inconsistent.
Similarly less supersymmetric D-branes constructed in this paper 
might be shown to be inconsistent by checking the Cardy condition.

\vs{.5cm}
\noindent
{\large\bf Acknowledgments}\\[.2cm]
I would like to thank M.\ R.\ Gaberdiel, N.\ Kim, F.-L.\ Lin, 
J.-H.\ Park, J.-T.\ Yee and P.\ Yi for helpful conversations and discussions.

\newcommand{\J}[4]{{\sl #1} {\bf #2} (#3) #4}
\newcommand{\andJ}[3]{{\bf #1} (#2) #3}
\newcommand{\AP}{Ann.\ Phys.\ (N.Y.)}
\newcommand{\MPL}{Mod.\ Phys.\ Lett.}
\newcommand{\NP}{Nucl.\ Phys.}
\newcommand{\PL}{Phys.\ Lett.}
\newcommand{\PR}{Phys.\ Rev.}
\newcommand{\PRL}{Phys.\ Rev.\ Lett.}
\newcommand{\PTP}{Prog.\ Theor.\ Phys.}
\newcommand{\hepth}[1]{{\tt hep-th/#1}}


\end{document}